\documentclass[12pt]{article}
\usepackage{psfig}
\usepackage{amsfonts}

\begin{document}
\thispagestyle{empty}
    \hfill{CTEQ-701}
\vskip 0.25cm
    \hfill{hep-ph/9701374}
\vskip 0.25cm
    \hfill{PSU/TH/177}
\vskip 0.25cm
    \hfill{CERN-TH/96-340}
\vskip 1cm
    \begin{center}
             {\Large \bf Diffractive Production of Jets and Weak Bosons,
                \\
             and Tests of Hard-Scattering Factorization
             }
    \end{center}
\vskip .5cm
\begin{center}
   Lyndon Alvero$^{a}$, John C. Collins$^{a,b}$, Juan Terron$^{c}$, and
   Jim Whitmore$^{a}$
\\
   $a)\ ${\it Physics Department, Pennsylvania State University}\\
      {\it 104 Davey Lab., University Park, PA 16802-6300,
      U.S.A.}\\[6pt]
   $b)\ ${\it CERN---TH division, CH-1211 Geneva 23, Switzerland.
   (Until 30 April 1997.)
   \\[6pt]
   $c)\ $ Universidad Aut\'onoma de Madrid,
          Departamenta de F\'isica Te\'orica,
          Madrid, Spain}
\end{center}
\vskip 1cm
    \centerline{24 January 1997}
\vskip .5cm
\centerline{\bf Abstract}
    We extract diffractive parton densities from diffractive,
    deep inelastic (DIS) $ep$
    data from the ZEUS experiment.
    Then we use these fits
    to predict the diffractive production of jets and
    of $W$'s and $Z$'s in $p\bar p$ collisions at the Tevatron.
    Although the DIS data
    require a hard quark density in the pomeron, we find fairly
    low rates for the Tevatron processes (a few percent of the inclusive
    cross section). This results from the combined effects of
    $Q^{2}$ evolution and of a normalization of the parton densities
    to the data.
    The calculated rates for $W$ production are generally
    consistent with the preliminary data from the Tevatron.
    However, the jet data from CDF with a ``Roman pot'' trigger
    are substantially lower than the results of our calculations;
    if confirmed, this would signal a breakdown of
    hard-scattering factorization.

\vfill
\noindent
CERN-TH/96-340\\
January 1997
\newpage

\section{Introduction}
\label{sec:intro}

In view of counterexamples \cite{nonfact} to the conjecture of
factorization \cite{ISorig} of hard processes in diffractive
scattering, it is important to test \cite{CTEQpom} factorization
experimentally.
In this paper, we present some results to this end.
Specifically, we present some preliminary fits to data on diffractive
deep inelastic scattering \cite{ZEUSdata}, and use these fits to
predict cross sections in hard diffractive processes in $p\bar p$
collisions, with the assumption of factorization.

We recall that diffractive events are characterized by a large rapidity
gap, a region in rapidity where no particles are produced. We are
concerned with the case where there is a hard scattering and the
gap occurs between the hard scattering and one of the beam
remnants. Such hard diffractive events have been observed in
deep inelastic scattering (DIS) experiments
\cite{firstdiff},
and are found to have a large rate: around 10\% of the inclusive
cross section.
Diffractive jet production in $p\bar p$ collisions was earlier
reported by the UA8 collaboration \cite{UA8}, but under somewhat
different kinematic conditions (larger $|t|$\footnote{
    By $t$ we mean the invariant momentum-transfer-squared from the
    diffracted hadron.
}).  There was also a
report of diffractive bottom production \cite{Eggert}.
Now, more diffractive data are being gathered from
a variety of lepto-hadronic \cite{ZEUSdata,H1,ZEUSphoto} and
hadronic processes \cite{CDF,CDF.Pot,CDF.ANL,D0.ANL,D0.1,D0.2},
but with substantially smaller
fractions in the case of the diffractive production of jets and weak
vector bosons in $p\bar p$ interactions.

The QCD-based model that we use in our calculations is the one due to
Ingelman and Schlein \cite{ISorig}, where diffractive scattering
is attributed to the exchange of a pomeron --- a colorless object
with vacuum quantum numbers. The pomeron is treated like a real
particle, and so a diffractive electron-proton collision is
considered to be due to an electron-pomeron collision.  Thus
hard cross sections involve a hard scattering coefficient (or
Wilson coefficient), a known pomeron-proton coupling, and parton
densities in the pomeron.  Similar remarks apply to diffractive
hadron-hadron scattering.

The parton densities in the pomeron can be extracted from
diffractive DIS $(F_{2})$ measurements.  Since the pomeron is
isosinglet and is its own charge conjugate, there is only a
single light quark density to measure; one does not have the
complications of separating the different flavors of quark that
one has in the case of the proton.  Scaling violations enable one
to determine the gluon density. The H1 collaboration has already
presented \cite{H1} a fit of this kind. This type of data
sufficiently determines the quark but, at present, it only weakly
constrains the gluon density in the pomeron.
The first experimental evidence for the gluon content of the
pomeron was found by the ZEUS collaboration  by combining their
results on the diffractive structure function in deep inelastic
scattering \cite{ZEUSdata} and their measurements of diffractive
inclusive jet photoproduction \cite{ZEUSphoto} under the
assumption of factorization between these two processes.
In this paper
we use each of several different fits that we have made to
diffractive $F_{2}$ data \cite{ZEUSdata} from ZEUS.

Our fits
to the DIS data
are made with full NLO calculations, but, given the
level of accuracy that is needed at present, we only use lowest
order QCD calculations for
the hadronic
processes. In the past, Ingelman and Schlein \cite{ISorig} and
Bruni and Ingelman \cite{BI} have made similar calculations for
one of the hadron-induced processes that we consider here.
Their results have provided a commonly used benchmark in the
phenomenology of these processes.
They provide a
choice of either `hard' or `soft' distributions of partons in the
pomeron, according to the $\beta \to 1$ behavior.\footnote{
   Here, $\beta $ is the fraction of the pomeron's momentum that is
   carried by the struck parton.
}
The hard distributions give larger diffractive cross sections.
At that time, there were no data to determine the distributions.
We will find that although the quark distributions preferred by
the DIS data are hard,
our cross sections are substantially below
those predicted by Bruni and Ingelman.  As we will see, the lower
cross sections occur for several reasons, particularly: a correct
normalization of the distributions to DIS data and the
incorporation of $Q^{2}$ evolution.

How well our predictions match up with the data in hadron-hadron
collisions will be a statement on the validity of factorization
of the diffractive hadronic cross sections. If there is good
agreement, comparison of our results with measured cross sections will also
provide a good test of the pomeron parton distribution fits from
diffractive DIS.  In addition, the results from diffractive jet
photoproduction \cite{ZEUSphoto} will provide greatly improved
constraints on the gluon content of the pomeron, and will
substantially improve the accuracy of our predictions.\footnote{
    As explained in Refs.\ \cite{nonfact,CTEQpom}, we consider it
    more likely that factorization is valid in diffractive
    deep inelastic and direct photoproduction than in diffractive
    hadron-induced processes.  Therefore, we prefer to determine
    the parton densities in the pomeron from deep inelastic
    scattering and from direct photoproduction, and then to treat
    the hadron-induced diffractive processes as providing tests
    of factorization.  Goulianos' proposal \cite{kgoul} to
    renormalize the pomeron flux in an energy dependent way could
    be regarded already as evidence that factorization is likely
    to break down.
    }

This paper is organized as follows.  In section \ref{sec:kin},
we present the formulae
used to calculate the various cross sections.  We also discuss the kinematics
and phase space cuts that were used.
In section \ref{sec:fits}, we present our fits to the diffractive
deep inelastic data.
Then in sections \ref{sec:VB.calcs} and \ref{sec:jet.calcs}, we
present and discuss the results obtained for vector boson
production and jet production, respectively.  We will find that
production rates for these diffractive events are no more than a
few percent, at most, of the nondiffractive ones. Finally, we
summarize our findings in section \ref{sec:concl}.

Other fits to the diffractive structure functions measured by H1
have been made by Gehrmann and Stirling \cite{GS} and by Kunszt
and Stirling \cite{KS}.
Golec-Biernat and Kwieci\'nski \cite{GK} assumed a
parameterization of the parton densities in the pomeron and found
it to be compatible with the H1 data on diffractive DIS.
Their quark densities are about 30\% smaller than ours, and
they required the momentum sum rule to be valid.
The new features of our work are a fit
to the ZEUS data, and a calculation of the cross sections for
diffractive jet and $W$ and $Z$ production.

%===================================================
\section{Kinematics and Cross Sections}
\label{sec:kin}

The diffractive processes that we consider here are
the production of $W$ and $Z$ bosons and of jets in
$p{\bar p}$ collisions.
In addition we consider $W$ production with explicit
consideration of the distribution of the final state leptons.
Schematically, these are
\begin{eqnarray}
p(p_{1})+{\bar p}(p_{2})&\to & (W\ {\rm or}\ Z) + {\bar p} + X, \nonumber \\
p(p_{1})+{\bar p}(p_{2})&\to & {\rm jet} + {\bar p} + X \nonumber \\
p(p_{1})+{\bar p}(p_{2})&\to & (W\to l + \nu ) + {\bar p} + X.
\label{gnricpro}
\end{eqnarray}
We take the pomeron to be emitted from the antiproton and the positive
$z$-axis to be along the antiproton's direction.  The center of mass
energy is $\sqrt {s}$, where $s=(p_{1}+p_{2})^{2}$.

%------------------------------------------
\subsection{Diffractive jet production}

Consider the diffractive cross section for the production of a
jet with rapidity $y$, in a hadron-hadron collision.
We will assume hard-scattering factorization
\cite{ISorig,CTEQpom}.
The lowest order
hard-scattering process is $2 \to  2$ at the parton level, and
results in a cross section of the form
\begin{equation}
{d\sigma ^{\rm jet} \over dy} =
\sum _{a,b}\int _{E_{T}^{\rm min}}^{E_{T}^{\rm max}}dE_{T} \, 2E_{T}
\int _{y'_{\rm min}}^{y'_{\rm max}}  dy'
\int _{x_{\mathbb P}^{\rm min}}^{x_{\mathbb P}^{\rm max}} dx_{\mathbb P}
\, f_{{\mathbb P}/{\bar p}}(x_{\mathbb P},\mu )
\,f_{a/p}(x_{a},\mu )
\,f_{b/{\mathbb P}}(x_{b},\mu )x_{a}x_{b}
\frac {d \hat{\sigma }^{\rm jet}_{ab}}{d \hat t} ,
\label{diffcs1}
\end{equation}
where the sum is over all the active parton (quark, antiquark and gluon)
flavors.  The integration variables are $E_{T}$, the transverse
energy of the jet, $y'$, the rapidity of the other jet,
and $x_{\mathbb P}$, the momentum fraction of the pomeron.  The
momentum fractions of the partons, relative to their parent
proton and pomeron are
\begin{eqnarray}
x_{a}&=&{E_{T} \over \sqrt {s}}(e^{-y}+e^{-y'}), \nonumber \\
x_{b}&=&{E_{T} \over \sqrt {s}x_{\mathbb P}}(e^{y}+e^{y'}).
\label{xpartons}
\end{eqnarray}
The functions $f_{a/p}(x_{a})$ and $f_{b/{\mathbb P}}(x_{b})$
are the distributions\footnote{
   These are number densities, not momentum densities.
}
of partons in the proton  and pomeron,
respectively,
while $f_{{\mathbb P}/{\bar p}}$ is the ``flux of pomerons in the
(anti)proton'', to be discussed below.
$d\hat{\sigma }^{\rm jet}_{ab}/d\hat{t}$ is the
partonic hard scattering coefficient and $\mu $ is the factorization
scale, which we set equal to $E_{T}$.  The
specific limits used for the integral in $x_{\mathbb P}$, as well as
those for the rapidity $y'$ and transverse energy $E_{T}$ will be
given later.

The diffractive cross section given by Eq.~(\ref{diffcs1}) has
the same structure as the factorized form of the corresponding
nondiffractive cross section, except for the pomeron flux factor
and the parton densities in the pomeron.
The same hard scattering coefficient and
nucleon parton distribution functions appear in both cross sections.

The pomeron flux factor,  $f_{{\mathbb P}/{\bar p}}$, is related to the
pomeron-proton coupling measured in proton-proton elastic
scattering.  The $t$ dependence (which we integrate over) and the
$x_{\mathbb P}$ dependence are thereby determined, but there is a
controversy as to the (constant) normalization needed to treat
the exchanged pomeron as if it were a particle.
Since the same normalization factor appears in all our cross
sections, its value will be irrelevant to our phenomenology.
Any change in the normalization factor is completely
compensated by changing the parton densities by an inverse
factor, and we obtain the parton densities from fitting a set of
data without any {\it a priori} expectations as to their
normalization.

However, the normalization does affect the question of whether
the momentum sum rule is obeyed by the parton densities in the
pomeron.  Since it is not at present understood whether the sum
rule is a theorem, this issue will not affect us.  The momentum
sum rule is {\em not} assumed in any of our fits.

There are two pomeron flux factors that are commonly used,
Ingelman-Schlein (IS) \cite{ISorig} and Donnachie-Landshoff (DL)
\cite{DLflux}.
Since the parton densities will be obtained
using the DL factor, we use the same factor in predicting
other cross sections.
The DL flux factor is given by
\begin{equation}
   f_{{\mathbb P}/{\bar p}}^{\rm DL}(x_{\mathbb P})=
   \int _{-\infty }^{0} dt \, \frac {9\beta _{0}^{2}}{4\pi ^{2}}
   \biggl[{4m_{p}^{2}-2.8t\over 4m_{p}^{2}-t}
      \biggl({1\over 1-t/0.7}\biggr)^{2}
   \biggr]^{2}x_{\mathbb P}^{1-2\alpha (t)},
\label{DLflux}
\end{equation}
where $m_{p}$ is the proton mass, $\beta _{0}\simeq 1.8\ {\rm GeV}^{-1}$ is the
pomeron-quark coupling and $\alpha (t)=1.085+0.25t$ is the pomeron trajectory.
The integral is over the invariant momentum transfer carried by
the pomeron, since in the present generation of measurements,
only the value of the longitudinal momentum of the pomeron is
measured.  Since the $t$ distribution is steeply falling, only
values of $|t|$ under about a ${\rm GeV}^{2}$ are significant.

The cross section given by Eq.~(\ref{diffcs1}) has contributions
from a range of subprocesses.  The indices $a,b$ labeling the
incoming partons range over the gluon and all the flavors of
quarks and antiquarks. The lowest order form of the partonic cross
section $d\hat{\sigma }^{\rm jet}_{ab}/d\hat{t}$ may be found in
\cite{eichetal}.

%-----------------------------------
\subsection{Diffractive $W$ and $Z$ production}

The cross section for the diffractive production of weak vector bosons,
is given by
\begin{equation}
    \sigma ^{W,Z} =
    \sum _{a,b}\int ^{x_{\mathbb P}^{\rm max}}_{x_{\mathbb P}^{\rm min}}
    dx_{\mathbb P}
    \int _{x_{b}^{\rm min}}^{1}dx_{b}f_{{\mathbb P}/{\bar p}}(x_{\mathbb P})
    f_{b/{\mathbb P}}(x_{b})
    f_{a/p}(x_{a})\tilde{C}_{ab}^{2}{1\over sx_{b}x_{{\mathbb P}}}
    \biggl[\sqrt {2}{\pi \over 3}G_{F}M_{VB}^{2}\biggr],
\label{vbincldfcs}
\end{equation}
where
$x_{b}$ is now the momentum fraction of
the parton from the pomeron, so that the momentum fraction of the
other parton (in the proton) is
$x_{a}=M_{VB}^{2}/x_{b}x_{{\mathbb P}}s$.  The minimum value of $x_{b}$ is
$x_{b}^{\rm min}=M_{VB}^{2}/x_{{\mathbb P}}s$.  Also,
$M_{VB}=M_{W} \ {\rm or} \ M_{Z}$ is the vector boson mass,
and $G_{F}$ is the Fermi constant.
For $W$ bosons, $\tilde{C}^{W}_{ab}=V_{ab}$,
the relevant Cabibbo-Kobayashi-Maskawa matrix element,
while for the $Z$ boson,
\begin{equation}
    \tilde{C}^{Z}_{ab}=
    \delta _{{\bar a}b}\biggl[{1\over 2}-2|e_{b}|\sin^{2}\theta _{W}
    +4|e_{b}|^{2}\sin^{4}\theta _{W}\biggr],
\label{Zconvfac}
\end{equation}
where $e_{b}$ is the fractional charge of parton $b$ and $\theta _{W}$ is the
Weinberg or weak-mixing angle.

%-----------------------------------
\subsection{Diffractive production of leptons from the $W$}

Since leptonic decays of $W$ bosons include an unobserved
neutrino, it is useful to compute the distribution of the
observed charged lepton.
The general formula for the distribution of leptons from $W$
production has the
same form as that for jet production, Eq.~(\ref{diffcs1}).
In the case of $W$ production, at lowest order there are no gluon
contributions.

For the specific process
$p+{\bar p}\to (W^{-}\to e + {\bar \nu }_{e})+{\bar p}+X$, we have
\begin{equation}
{d\hat{\sigma }^{\rm lep}_{ab}\over d\hat{t}}\simeq
{G_{F}^{2}\over 6M_{W}\Gamma _{W}}V_{ab}^{2}
\delta (x_{a}x_{b}s-M_{W}^{2})\hat{u}^{2},
\label{wmpartcs}
\end{equation}
where $M_{W} (\Gamma _{W})$ is the mass (width) of the $W$
boson, $V_{ab}$ is the
Cabibbo-Kobayashi-Maskawa matrix element and
\begin{equation}
\hat{u}=-x_{b}x_{\mathbb P}\sqrt {s}E_{T}e^{-y}.
\label{defuhat}
\end{equation}
We used the narrow width approximation in Eq.\ (\ref{wmpartcs}).
Using Eq.~(\ref{wmpartcs}) in Eq.~(\ref{diffcs1}), one obtains
\begin{equation}
    {d\sigma ^{\rm lep} \over dy}=
    \sum _{a,b}\int _{x_{\mathbb P}^{\rm min}}^{x_{\mathbb P}^{\rm max}}
        {dx_{\mathbb P}\over x_{\mathbb P}}
    \int _{E_{T}^{\rm min}}^{E_{T}^{\rm max}}  dE_{T}
    f_{{\mathbb P}/{\bar p}}(x_{\mathbb P})
    f_{b/{\mathbb P}}(x_{b})f_{a/p}(x_{a})V_{ab}^{2}
    \left[{\hat{u}^{2}G_{F}^{2}\over 6s\Gamma _{W}[(M_{W}/2E_{T})^{2}-1]^{1/2}}
    \right],
\label{wmdiffcs}
\end{equation}
where
$x_{a}, x_{b}$ are now given by
\begin{eqnarray}
x_{a}&=& \frac {M_{W} e^{-y}}{\sqrt s}
     \left[\frac {M_{W}}{2E_{T}}
        + \sqrt { \left(\frac {M_{W}}{2E_{T}}\right)^{2}-1 }
     \right],
\nonumber \\
x_{b}&=&{M_{W}^{2}\over s}{1\over x_{a}x_{\mathbb P}}.
\label{wmxpart}
\end{eqnarray}
We have suppressed the scale dependence of
the functions $f_{i/j}$ in Eqs.~(\ref{vbincldfcs})
and (\ref{wmdiffcs}); we set the scale
equal to the vector boson mass.
A similar equation may be
obtained for the $W^{+}$ cross section.

%-----------------------------------
\subsection{Inclusive cross sections}

Since we are particularly interested in the percentage of events
that are diffractive, we also need to calculate the inclusive
cross sections, that is, the ones without the diffractive
requirement on the final state.
The analog to Eq.~(\ref{diffcs1}) for the
inclusive cross section for jet production is the
standard formula
\begin{equation}
    {d\sigma ^{\rm jet,\ incl} \over dy}=
    \sum _{a,b}\int _{E_{T}^{\rm min}}^{E_{T}^{\rm max}} dE_{T}2E_{T}
    \int _{y'_{\rm min}}^{y'_{\rm max}}dy'
    f_{a/p}(x_{a},\mu )f_{b/{\bar p}}(x_{b},\mu )x_{a}x_{b}
    \frac {d\hat{\sigma }^{\rm jet}_{ab}}{d \hat t},
\label{nondiffcs1}
\end{equation}
where $x_{a}$ is given in Eq.~(\ref{xpartons}) while $x_{b}$ is now
$x_{b}=(e^{y}+e^{y'}) E_{T}/ \sqrt {s}$.

For
the leptons from
$W^{-}$ production, the inclusive version of Eq.~(\ref{wmdiffcs}) is
\begin{equation}
{d\sigma ^{\rm lep,\ incl} \over dy} =\sum _{a,b}
\int _{E_{T}^{\rm min}}^{E_{T}^{\rm max}} dE_{T}
f_{b/{\bar p}}(x_{b},\mu )f_{a/p}(x_{a},\mu )V_{ab}^{2}
\biggl[{\hat{u}^{2}G_{F}^{2}\over 6s\Gamma _{W}[(M_{W}/2E_{T})^{2}-1]^{1/2}}
\biggr],
\label{wmndiffcs}
\end{equation}
with a similar equation for $W^{+}$ production.  In Eq.~(\ref{wmndiffcs}),
$\hat{u}=-x_{b}\sqrt {s}E_{T}e^{-y}$,
$x_{a}$ is as defined in Eq.~(\ref{wmxpart}) while $x_{b}$ is now given by
$x_{b}=M_{W}^{2}/x_{a}s$.

The analog to Eq.~(\ref{vbincldfcs}) for the inclusive total cross section
for vector boson production is
\begin{equation}
\sigma ^{W,Z}=\sum _{a,b}\int _{x_{b}^{\rm min}}^{1}dx_{b}
f_{b/{\bar p}}(x_{b})f_{a/p}(x_{a})\tilde{C}_{ab}^{2}{1\over sx_{b}}
\biggl[\sqrt {2}{\pi \over 3}G_{F}M_{VB}^{2}\biggr],
\label{vbinclndfcs}
\end{equation}
where $\tilde{C}_{ab}$ and $M_{VB}$ are as
defined above, $x_{b}^{\rm min}=M_{VB}^{2}/s$
and $x_{a}=M_{VB}^{2}/x_{b}s$.

%========================================================
\section{Partons in the Pomeron}
\label{sec:fits}

We have made five fits of parton densities in the pomeron from the
ZEUS data in Ref.\ \cite{ZEUSdata} for the diffractive structure
function $F_{2}^{D}$.  Diffractive structure functions are related to
the differential cross section for the process $e + p \to  e + p + X$:
\begin{equation}
   \frac {d^{4}\sigma _{\rm diff}}{d\beta  dQ^{2} dx_{\mathbb P} dt}
   = \frac {2\pi \alpha ^{2}}{\beta Q^{4}}
     \left\{
        \left[ 1 + (1-y)^{2} \right] F_{2}^{D(4)}
        - y^{2} F_{L}^{D(4)}
     \right\} ,
\end{equation}
where corrections due to $Z^{0}$ exchange and due to radiative
corrections have been ignored.  Here $x_{\mathbb P}$ and $t$ are the
same as in the previous sections, $Q^{2}$ and $y$ are the usual DIS
variables, and $\beta =x_{\rm bj}/x_{\mathbb P}$, with $x_{\rm bj}$ being the
usual Bjorken scaling variable of DIS.  The diffractive structure
function $F_{2}^{D(4)}(\beta ,Q^{2},x_{\mathbb P},t)$ is assumed to obey Regge
factorization, so that it is written as a pomeron flux factor
times a pomeron structure function:
\begin{equation}
    F_{2}^{D(4)}(\beta ,Q^{2},x_{\mathbb P},t)
    = f_{\mathbb P}(x_{\mathbb P},t)
      F_{2}^{\mathbb P}(\beta ,Q^{2}).
\end{equation}
The momentum transfer $t$ is not measured, so the data are
actually for the structure function integrated over $t$:
\begin{equation}
    F_{2}^{D(3)}(\beta ,Q^{2},x_{\mathbb P})
    = \int _{-\infty }^{0} dt
      F_{2}^{D(4)}(\beta ,Q^{2},x_{\mathbb P},t) .
\end{equation}
Furthermore, the actual fits are to $\tilde F_{2}^{D}(\beta ,Q^{2})$, which is
obtained by integrating $F_{2}^{D(3)}(\beta ,Q^{2},x_{\mathbb P})$
over the measured range of $x_{\mathbb P}$, $6.3 \cdot 10^{-4} <
x_{\mathbb P} < 10^{-2}$, using the {\it fitted} $x_{\mathbb P}$
dependence \cite{ZEUSdata}. That procedure, as noted in
\cite{ZEUSdata}, assumes that a universal $x_{\mathbb P}$
dependence holds in {\it all} regions of $\beta $ and $Q^{2}$.

Hard scattering factorization gives $F_{2}^{\mathbb P}$ in terms of
parton densities and hard scattering coefficients in the usual
fashion:
\begin{equation}
    F_{2}^{\mathbb P}(\beta ,Q^{2})
    = \sum _{a} e_{a}^{2} \beta  f_{a/\mathbb P}(\beta )
      + \mbox{NLO corrections} .
\end{equation}
Since the outgoing proton is not detected, the data include
contributions where the proton is excited to a state that
escapes down the beam-pipe and thus misses
the detector.  Excited states up to about 4 GeV pass the
diffraction selection cuts, and the experimenters estimate that
there is a contribution of $(15\pm 10)\%$ to the measured
diffractive $F_{2}$ from such ``double-dissociative'' events.
This point is significant when we compare predictions obtained using
our fits to data where the diffracted proton is detected, as in
Sect.\ \ref{sec:jet.calcs}.

Each of our fits is represented by a
parameterization of the initial distributions
at $Q_{0}^{2} = 4 \, {\rm GeV}^{2}$ for the $u$,
$\bar u$, $d$, and $\bar d$ quarks and for the gluon.
The other quark distributions are assumed to be zero at this
scale.  The fits
were made with NLO calculations
(with full evolution and with the number of flavors set equal to
4) and with the pomeron flux factor chosen to be that
of Donnachie and Landshoff.\footnote{
    The flux factor is a common factor in all the cross
    sections we compute, and we do not assume a momentum sum rule
    for the parton densities in the pomeron.
    Therefore, the choice of flux factor does not affect our
    predictions for the hadron-induced processes provided only
    that the pomeron trajectory function $\alpha (t)$ is correct.
}
The program used to perform the evolution was that of CTEQ
\cite{CTEQ}.

\begin{figure}
\centerline{
\psfig{file=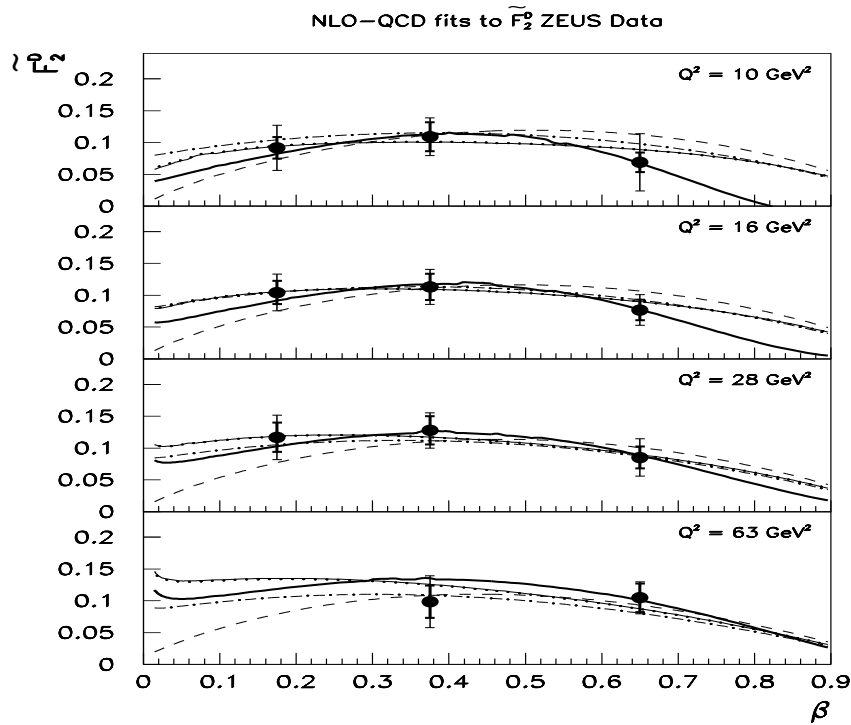,height=4in,width=6in,clip=}}
\caption{\sf
        The $\beta $ dependence of the diffractive structure
        function $\tilde F_{2}^{D}$ measured by ZEUS, together with our
        fits.
        The inner, thick error bars represent the
        statistical errors on the data, and the outer, thin error
        bars represent the systematic and statistical errors
        added in quadrature.
        Fit A is represented by the dashed line, fit
        B by the thin solid line, fit C by the dot-dashed line,
        fit D by the dotted line, and fit SG by the thick solid
        line.  Note that fits B and D are essentially
        indistinguishable. }
\label{DIS.beta}
\end{figure}

\begin{figure}
\centerline{
\psfig{file=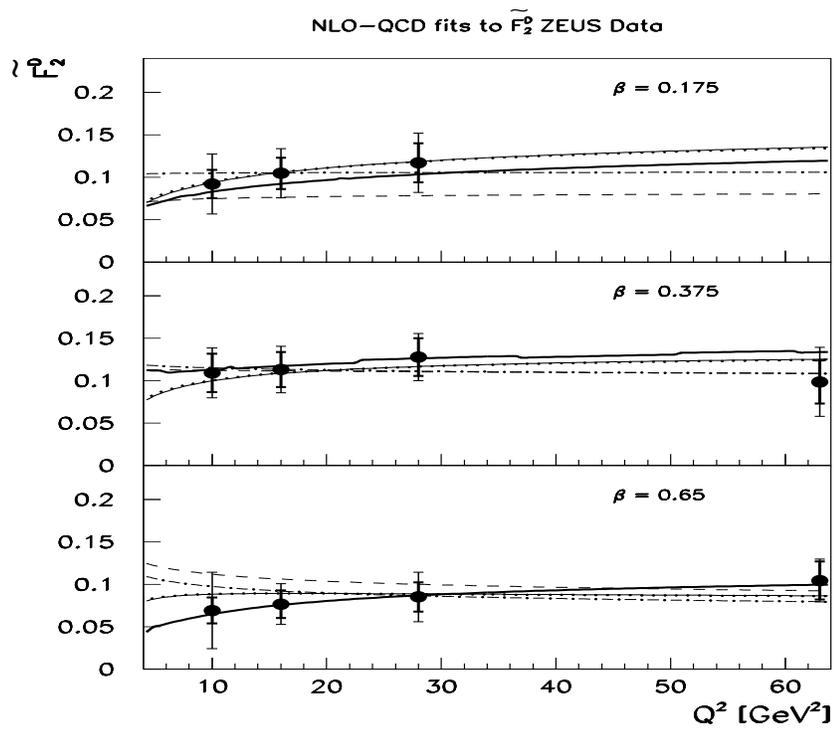,height=4in,width=6in,clip=}}
\caption{\sf
        The $Q^{2}$ dependence of the diffractive structure
        function $\tilde F_{2}^{D}$ measured by ZEUS, together with our
        fits.
        The description of the
        lines and error bars is the same as in Fig.\ \ref{DIS.beta}.
}
\label{DIS.Q2}
\end{figure}

Four of the fits, labeled ``A'', ``B'', ``C'' and ``D'', use
conventional shapes for the initial distributions.  The final fit
has a gluon distribution that is peaked near $\beta =1$, as suggested
by the fit \cite{H1} exhibited by the H1 collaboration; we call
this our ``singular gluon'' fit, SG.
We show our fits in Figs.\ \ref{DIS.beta} and \ref{DIS.Q2}.

Our first fit, A, has as its initial distribution a hard quark
distribution (proportional to $\beta (1-\beta )$) with no glue, and hence
one adjustable parameter, whose value is the result of the fit:
\begin{eqnarray}
    \beta f^{A}_{q/{\mathbb P}}(\beta ,Q_{0}^{2}) &=&
    0.585 \, \beta  (1-\beta ) ,
\nonumber \\
    \beta f^{A}_{g/{\mathbb P}}(\beta ,Q_{0}^{2}) &=& 0 .
\label{fitA}
\end{eqnarray}
This fit gives the dashed line in the figures.  It represents the
shape of the diffractive structure function $\tilde F_{2}^{D}$ moderately
well, but there are noticeable deviations, both as a function of
$Q^{2}$ and of $\beta $.

For the $Q^{2}$-dependence, Fig.\ \ref{DIS.Q2}, we see that fit A
results in an $\tilde F_{2}^{D}$
that at large $\beta $ decreases with $Q^{2}$,
whereas the data show a tendency to increase.  At small $\beta $,
both the data and the fit rise with $Q^{2}$, but the data rise more
rapidly.  This suggests adding in an initial gluon distribution,
whose effect is to make the quark distributions rise. As to the
$\beta $-dependence, we see from Fig.\ \ref{DIS.beta}, that the fit
gives an $\tilde F_{2}^{D}$
that is rather higher than the data at large $\beta $
and rather lower at small $\beta $.  This suggests that an admixture
of soft quarks, more peaked at small $\beta $ would improve the fit
noticeably.

So we try each of these additions in turn, and then together.
First, in fit B we allow a hard gluon in the initial
distribution, with the result:
\begin{eqnarray}
    \beta f^{B}_{q/{\mathbb P}}(\beta ,Q_{0}^{2}) &=&
    0.516 \, \beta  (1-\beta ) ,
\nonumber \\
    \beta f^{B}_{g/{\mathbb P}}(\beta ,Q_{0}^{2}) &=&
    12.28 \, \beta  (1-\beta  ) .
\label{fitB}
\end{eqnarray}
Since the gluons only affect deep inelastic scattering in
next-to-leading order (NLO), a very large gluon distribution is
needed to produce a substantial effect on $\tilde F_{2}^{D}$.  The resulting
$Q^{2}$ evolution has a much more satisfactory shape, as can be seen
from the thin solid line in Fig.\ \ref{DIS.Q2}.  In addition, the
strong gluon distribution has the effect of biasing the evolved
quark distribution towards smaller $\beta $.  Thus the shape of
$\tilde F_{2}^{D}$
as a function of $\beta $ is also improved, Fig.\
\ref{DIS.beta}.\footnote{
    Remember that the initial $Q^{2}$ is $Q_{0}^{2}=4 \, {\rm GeV}^{2}$, so
    that evolution already has a significant effect at the lowest
    value of $Q^{2}$ for which we use data.
}

Next, in fit C, represented by the dot-dashed line, we see the
effect of adding a soft term to the quark distribution without
allowing an initial gluon distribution.  In general we expect a
soft quark term, since Regge theory predicts that a quark
distribution at small $\beta $ behaves approximately as $\beta f \sim 1$.
We
find the best fit of this kind to be:
\begin{eqnarray}
    \beta f^{C}_{q/{\mathbb P}}(\beta ,Q_{0}^{2}) &=&
                0.470 \, \beta  (1-\beta )
                + 0.080 \, (1-\beta )^{2},
\nonumber \\
    \beta f^{C}_{g/{\mathbb P}}(\beta ,Q_{0}^{2}) &=& 0 .
\label{fitC}
\end{eqnarray}
The normalization of the hard quark term is reduced by about
20\%, compared with fit A, and the added soft term clearly
improves the shape of the $\beta $ distributions, in Fig.\
\ref{DIS.beta}.  It should be clear that at moderate and large
values of $\beta $, above about 0.2, there is a dominant hard quark
term.  Moreover a soft term, at small $\beta $ is needed, no matter
whether it is intrinsic to the quark distribution or whether it
is generated dynamically, from evolution controlled by the gluon
distribution. As is to be expected, the soft quark term does
nothing to improve the $Q^{2}$ dependence in Fig.\ \ref{DIS.Q2}.

We next try both a soft quark term and a hard gluon term.  This
results in fit D
(shown as the dotted line in the figures):
\begin{eqnarray}
    \beta f^{D}_{q/{\mathbb P}}(\beta ,Q_{0}^{2}) &=&
        0.512 \, \beta  (1-\beta )
        + 0.005 \, (1-\beta )^{2},
\nonumber \\
    \beta f^{D}_{g/{\mathbb P}}(\beta ,Q_{0}^{2}) &=&
        11.65 \, \beta  (1-\beta  ) .
\label{fitD}
\end{eqnarray}
This is indistinguishable from fit B, particularly given the
errors.  The important fact is that if the pomeron were quark
dominated, QCD predicts that $\tilde F_{2}^{D}$
decreases with $Q^{2}$ at large
$\beta $.  The only way of undoing, or even reversing, this evolution
is to have a large initial amount of glue.  No extra soft quark term is
required by the data.

Given the number of points and their errors, we can conclude that
\begin{itemize}

\item Hard quark and gluon distributions are preferred.

\item The normalization of the quark distribution is reasonably
    well determined.

\item Probably a substantial gluon distribution is preferred.

\end{itemize}
However, even the pure quark fit, A, gives a satisfactory $\chi ^{2}$
--- see Eq.\ (\ref{chi2}) below.  It should be remembered,
however,  that $\chi ^{2}$ is not the only measure of goodness-of-fit.
On the other hand, the error bars in the lowest plot in Fig.\
\ref{DIS.Q2} indicate that reduction of the systematic errors in
the measurement of $\partial \tilde F_{2}^{D}/\partial Q^{2}$ is urgently
needed to improve the determination of the gluon distribution.

Finally, we recall that the H1 collaboration has shown fits to
their data with a gluon distribution that is very peaked close to
$\beta =1$.  So we have tried a similar fit.
The result, fit SG
(shown as the thick solid line in the figures), is:
\begin{eqnarray}
    \beta f^{SG}_{q/{\mathbb P}}(\beta ,Q_{0}^{2}) &=&
        0.354 \, \beta  (1-\beta ) ,
\nonumber \\
    \beta f^{SG}_{g/{\mathbb P}}(\beta ,Q_{0}^{2}) &=&
        70.756 \, \beta ^{8}  (1-\beta )^{0.3}.
\label{fitSG}
\end{eqnarray}
The exponents in the gluon density of Eq.~(\ref{fitSG}) were
chosen to try and match the singular gluon parameterization
derived by H1 \cite{H1}; the ZEUS data, at least at this time,
are not able to determine definite values for these exponents.  To
avoid possible numerical problems with the evolution code we have
chosen a shape for the gluon distribution that has a peak close
to $\beta =1$ but that is not actually singular there.

The singular fit results in an improved $\chi ^{2}$ --- see Eq.\
(\ref{chi2}) below --- presumably because the resulting
$\tilde F_{2}^{D}$
increases more rapidly with $Q^{2}$ at large $\beta $.  Given the size of
the systematic error bars in the relevant plot, the bottom one of
Fig.\ \ref{DIS.Q2}, this reinforces the need for improving the
systematic errors on $\partial \tilde F_{2}^{D}/\partial Q^{2}$.

In each of our fits, the parameters of the fit are the
nonzero coefficients, with the exponents being held fixed.  The
parameters are determined by minimizing the $\chi ^{2}$ between the
theory and the data.  The systematic and statistical errors have
been added in quadrature, although that is not an ideal
procedure.
All the fits give a good $\chi ^{2}$
and the fits with the large gluon are somewhat preferred,
particularly the singular gluon fit:
\begin{equation}
   \begin{tabular}{r|c|c|c|c|c}
                  & Fit A  & Fit B  & Fit C  & Fit D & Fit SG
   \\ \hline
      $\chi ^{2}/\mbox{d.o.f}$
                  & 5.7/10 & 1.7/9  & 2.5/9  & 1.7/8 & 1.2/9
   \\
      Statistical $\chi ^{2}/\mbox{d.o.f}$
                  & 18/10  & 3.8/9  & 6.0/9  & 3.8/8 & 3.1/9
   \end{tabular}
\label{chi2}
\end{equation}
The fact that the $\chi ^{2}$ per degree of freedom for all these fits
is much less than unity presumably indicates that the systematic
errors have not been treated correctly, for example, as regards
point-to-point correlations.
However, even if only the statistical errors are taken into
account, 4 of the 5 fits have a worryingly low $\chi ^{2}$ per degree of
freedom, as can be seen from the last line.  There is only a few
per cent probability of getting fits this good, even if the theory
is exactly correct.  We take this as an indication that one
should look more carefully at the computation of the errors on
the structure functions.

Independently of the quantification of the errors, the most
important feature from our point of view is that the fits give
rather stable normalizations to the quark distributions. The second important
feature is that a very large gluon distribution is preferred,
perhaps strongly peaked at large $\beta $. This is in agreement with
the results presented by the H1 collaboration \cite{H1}.
The momentum sums
$\sum _{a}\int _{0}^{1} d\beta  \beta  f_{a/\mathbb P}(\beta )$
are given in the following table:
\begin{equation}
   \begin{tabular}{r||l|l|l|l|l}
               & Fit A & Fit B & Fit C & Fit D & Fit SG
   \\ \hline\hline
        Quarks & 0.39  & 0.34 & 0.42   & 0.35  & 0.24
   \\
        Gluons & 0     & 2.05 & 0      & 1.94  & 3.57
   \\ \hline
        Total  & 0.39  & 2.39 & 0.42   & 2.29  & 3.81
   \end{tabular}
\end{equation}
Although the allowed gluon distributions cover a wide range, the
overall normalization of the quarks is determined to about
$\pm 10\%$, except in the case of the singular gluon fit, where the
quarks are brought down by about 30\%.

%===================================================
\section{Numerical Calculations of $W$ and $Z$ Production}
\label{sec:VB.calcs}

For the calculations in this section, the factorization scale in the parton
distributions was set to $M_{VB}$.  The values of the electroweak parameters
which appear in the various formulae were taken from the particle data
handbook \cite{PDG},
and we use only four flavors (u, d, s, c)
in the weak mixing matrix,
with the Cabibbo angle $\theta _{C} = 0.2269$.

%-----------------------------------
\subsection{Comparison to previous calculations}

Bruni and Ingelman \cite{BI} computed diffractive $W/Z$ cross sections
up to
${\cal O}(\alpha _{s})$, i.e., including gluon contributions.  These
calculations neglected any $Q^{2}$ evolution of the parton distributions in
the pomeron.  At $\sqrt {s}=1800 \ {\rm GeV}$, they obtained the
following
diffractive fractions ($R=\sigma ^{\rm diff} /\sigma ^{\rm incl}$):
$R_{W^{+}+W^{-}}\simeq 17\%$ and
$R_{Z}\simeq 15\%$ for total $W$ and $Z$ production, respectively.
These rates are substantially larger than the few percent quoted in preliminary
CDF results \cite{CDF,CDF.ANL}.

As we will now explain, when one uses evolved pomeron parton densities
from the above fits to data from the ZEUS experiment, one obtains
substantially smaller rates than the Bruni-Ingelman ones.
To understand these small rates, we first
verify that we can reproduce the Bruni-Ingelman results.
For these we used
their unevolved hard quark distribution in
the pomeron (given by their Eq.~(4)),
the same cut on $x_{{\mathbb P}}$: $x_{{\mathbb P}}^{\rm max}=0.1$, the
EHLQ1 parton distributions in the proton
and the Ingelman-Schlein (IS) flux factor:\footnote{
    Note that since our purpose in using the IS flux is to
    compare our results with the Bruni-Ingelman calculations, we
    have used a pomeron intercept of unity instead of the more
    accurate value used in the DL factor Eq.\ (\ref{DLflux}).
}
\begin{equation}
f_{{\mathbb P}/p}^{\rm IS}(x_{{\mathbb P}})=\int dt {1\over 2.3x_{{\mathbb P}}}
\biggl(6.38e^{8t}+0.424e^{3t}\biggr).
\label{ISflux}
\end{equation}
Next, we evolved their pomeron parton distributions and
recalculated the
cross sections.
Finally, to provide our best estimates of the rates,
we repeated the calculations using CTEQ3M for
the parton densities in
the proton/antiproton and using our fits
for the parton densities in the pomeron, all with proper
evolution.\footnote{
   We evolved the BI distribution
   from $Q_{0}^{2}=5 \, {\rm GeV}^{2}$ (as with the EHLQ1 distributions),
   while our fits were evolved
   from $Q_{0}^{2}=4 \, {\rm GeV}^{2}$.}
All the inclusive cross sections were calculated using
Eqs.~(\ref{vbincldfcs}) and (\ref{vbinclndfcs}).
The results we obtained are summarized in Tables
\ref{table:inclWZ} to \ref{table:diffWZ0.01}.

First, in Table \ref{table:inclWZ}, we show
the {\em inclusive}\footnote{
    I.e., diffractive plus non-diffractive.
}
cross section, $\sigma ^{\rm incl}$,
which will give the
denominator for the fraction of the cross section which is
diffractive.  Our results are shown in the last column.  These
are 20\% to 30\% higher than those of Bruni and Ingelman (column
2).  We have verified (column 3) that this increase is due to the
use of the more up-to-date CTEQ3M densities in the proton instead
of the EHLQ1 densities used by Bruni and Ingelman.
Given the current accuracy of the diffractive data,
we did not bother to make NLO calculations; generally the effect
would be to increase the cross sections by some tens of percent.

\begin{table}
\begin{center}
   \begin{tabular}{||c|c|c|c||}
   \hline
   &\multicolumn{1}{|c|}{Ref.\cite{BI} LO} & & \\
   &\multicolumn{1}{c}{EHLQ1} & \multicolumn{1}{|c|}{EHLQ1} &
   \multicolumn{1}{c||}{CTEQ3M} \\ \hline
   $W^{+}+W^{-}$ & 14000 & 14332 & 18150   \\ \hline
   $Z$     &  4400 &  4407 &  5383   \\ \hline
   \end{tabular}
\end{center}
\caption{\sf Inclusive cross sections $\sigma ^{W,Z\,\rm incl}$ (pb) for
         weak vector boson production.
}
\label{table:inclWZ}
\end{table}

The diffractive cross sections $\sigma ^{W,Z\, \rm diff}$ are
shown in Tables
\ref{table:diffWZ0.1} and \ref{table:diffWZ0.01}.
In the columns labeled `BI', we used the Bruni-Ingelman parton
density in the pomeron and the EHLQ1 parton densities in the
proton.
In the other columns we used our fits for the parton densities in
the pomeron together with the CTEQ3M parton distributions in the
proton.
First,
we use the same cut $x_{{\mathbb P}}^{\rm max} = 0.1$ that was used by
Bruni and Ingelman, to produce Table \ref{table:diffWZ0.1}.
However, this allows $x_{\mathbb P}$ to be rather larger
than where pomeron exchange is expected to dominate.  So we also
made calculations with $x_{{\mathbb P}}^{\rm max} = 0.01$,
for which the results
are shown in Table \ref{table:diffWZ0.01}.

In column 3 of Table \ref{table:diffWZ0.1} we show our results
when we use the same unevolved parton densities as Bruni and
Ingelman; we agree with their cross sections.  Then we repeat the
calculations but with
correctly evolved parton densities in the pomeron, with the
Bruni-Ingelman formula being used as the initial data for the
evolution at $Q_{0}^{2}=5 \, {\rm GeV}^{2}$.
We see
that this leads to about a 30\% reduction in the cross section.
The diffractive fraction obtained from the evolved BI pomeron
parton distribution, using column 3 of Table \ref{table:inclWZ}
for $\sigma ^{W,Z\, \rm incl}$,
is about 14\% for $W$ production, compared with the
19\% that is obtained using the unevolved BI pomeron
distributions.
The corresponding percentages for $Z$ production are a little
smaller: 12\% and 17\%.

\begin{table}
\begin{center}
   \begin{tabular}{||c|c|c|c|c|c||}
   \hline
   Pomeron: &  BI\cite{BI} &  Our BI   &   BI    & Fit A   & Fit D \\
            &  unevolved   & unevolved & evolved & evolved & evolved \\
   Proton:  &    EHLQ1     &   EHLQ1   &  EHLQ1  & CTEQ3M  & CTEQ3M  \\
   \hline
   $W^{+}+W^{-}$ & 2800    & 2768      & 2025    & 518     & 844 \\ \hline
   $Z$      &       760    &  738      &  520    &  133    &  204 \\ \hline
\end{tabular}
\end{center}
\caption{\sf Diffractive cross section $\sigma ^{W,Z\, \rm diff} $ (pb)
    for weak vector boson
    production, with $x_{\mathbb P}^{\rm max} = 0.1$.}
\label{table:diffWZ0.1}
\end{table}

\begin{table}
\begin{center}
   \begin{tabular}{||c|c|c|c||}
   \hline
   Pomeron:  &   BI (evolved)   & Fit A (evolved) & Fit D (evolved) \\
   Proton:   &      EHLQ1       &      CTEQ3M     &  CTEQ3M  \\
   \hline
   $W^{+}+W^{-}$ &   52.3       &       12.8        & 13.9\\ \hline
   $Z$        &       6.6       &        1.6        &  1.6 \\ \hline
\end{tabular}
\end{center}
\caption{\sf Diffractive cross section $\sigma ^{W,Z\, \rm diff} $ (pb)
    for weak vector boson
    production, but now with $x_{\mathbb P}^{\rm max} = 0.01$.}
\label{table:diffWZ0.01}
\end{table}

In the last two columns of Tables \ref{table:diffWZ0.1} and
\ref{table:diffWZ0.01} we present the results when two of our
fits to ZEUS data are used. Fit A is the one with a simple hard
quark distribution and no glue as the initial values, while fit D,
together with fit B, has the smallest $\chi ^{2}$.
Although this latter fit has a very large gluon
distribution which carries about seven times as much momentum as
the quarks, the preference for a large amount of glue is
mild---see the last two columns of (\ref{chi2}).
Furthermore, at $Q_{0}^{2}=4 \, {\rm GeV}^{2}$, the initial quark
distribution remains mostly unchanged whether there are zero
gluons or, conversely, the gluons are unconstrained.

The cross sections resulting from fit A (in column 5 of Table
\ref{table:diffWZ0.1}) are about
25\% of the evolved BI cross sections.  The
diffractive fractions obtained from
this fit, using the CTEQ3M entries in Table \ref{table:inclWZ},
are 2.9\% (2.5\%) for $W\ (Z)$ production,
as shown in Table \ref{table:fractions}.

In fit D, there is an enormous amount of glue initially, and this
significantly affects the evolution of the quarks in the pomeron
from $Q_{0}^{2}=4 \, {\rm GeV}^{2}$ to the vector boson mass.  The result is
an increase in the cross section, as can be seen in column 6 of
Table \ref{table:diffWZ0.1}.  This happens even though the
initial quark distribution is smaller than in fit A. Even so, the
cross sections are still smaller, by a factor of about 2, than
the ones from evolved BI pomeron parton distributions.  The rates
from fit D are 4.7\% (3.8\%) for $W\ (Z)$ production.
These rates are somewhat lower than those obtained by Kunszt and
Stirling \cite{KS}, who mostly used quark distributions
that at large $\beta $ fall less steeply than ours.

The data \cite{ZEUSdata} from which our fits were extracted
used a conservative cut
on the pomeron momentum, $x_{{\mathbb P}}^{\rm max}=0.01$.
The pomeron flux factor allows for the $x_{\mathbb P}$ dependence, but
to ensure maximum compatibility with the ZEUS data without the
assumption of standard Regge behavior, the same cut should be
applied to the cross sections in hadron-hadron collisions.  This
results in the cross sections in Table \ref{table:diffWZ0.01},
which therefore represent our most accurate prediction of
diffractive $W$ and $Z$ production, given only the assumption of
hard scattering factorization,
{\em which of course we wish to test}.
Notice that with this cut the diffractive
cross sections are over an order of magnitude smaller
than with
$x_{{\mathbb P}}^{\rm max}=0.1$.
The percentages obtained with this cut on $x_{{\mathbb P}}$
for $W\ (Z)$ production are 0.07\% (0.03\%) and 0.08\% (0.03\%)
for fits A and D, respectively,
as shown in Table \ref{table:fractions}
The large reduction is presumably due to
the fact that we are not far from an effective kinematic limit:
the cut on $x_{\mathbb P}$ gives a maximum proton-pomeron energy of
180 GeV, and partons typically do not carry the whole of the
energy of their parent hadrons.

\begin{table}
    \begin{tabular}{||c|c||c|c||c|c||}
        \hline
                & BI (unevolved) & Fit A & Fit D & Fit A & Fit D \\
                & $x_{\mathbb P}^{\rm max} = 0.1$
                    & $x_{\mathbb P}^{\rm max} = 0.1$
                        & $x_{\mathbb P}^{\rm max} = 0.1$
                            & $x_{\mathbb P}^{\rm max} = 0.01$
                               & $x_{\mathbb P}^{\rm max} = 0.01$
        \\
        \hline
        $W^{+}+W^{-}$ & 19\% & 2.9\% & 4.7\% & 0.07\% & 0.08\% \\
        \hline
        $Z$     & 17\% & 2.5\% & 3.8\% & 0.03\% & 0.03\% \\
        \hline
    \end{tabular}
\caption{\sf Diffractive fractions}
\label{table:fractions}
\end{table}

%----------------
\subsection{Why are the fractions smaller than from BI?}

Although the data used in our fits support a ``hard'' quark
distribution\footnote{
    This agrees with the H1 results \cite{H1}.
}
in the pomeron, we predict that the diffractive
$W$ and $Z$ cross sections are
much smaller, by a factor of 3 to 5, than those predicted by
Bruni and Ingelman, who also used hard quark distributions.  This
factor arises as an accumulation of several modest factors that
all change the ratios in the same direction.
\begin{itemize}

\item
    A factor 0.8 because of the larger inclusive cross sections
    when one uses CTEQ3M instead of the obsolete EHLQ1
    distributions in the proton.

\item
    A factor 0.7 for the effect of the evolution of the parton
    densities in the pomeron.

\item
    A factor 0.7 for the use of the Donnachie-Landshoff flux
    factor instead of the Ingelman-Schlein flux factor, when the
    momentum sum is kept fixed.

\item
    A factor 0.5 because the data indicate that the quarks
    give a contribution to the momentum sum of 0.5 (with the DL
    normalization), instead of unity as assumed by Bruni and
    Ingelman.

\end{itemize}

%-----------------------
\subsection{Lepton distributions for $W$ production at the Tevatron}

In this section, we present our results for $W$ production, but
now with cuts on the
emitted lepton $l$.  Specifically, we calculate the electron's (or positron's)
rapidity $(y)$
distribution from Eq.~(\ref{wmdiffcs}) for the diffractive process, and
Eq.~(\ref{wmndiffcs}) for the inclusive one.
For the parton distributions in the pomeron, we use
our five fits, Eqs.~(\ref{fitA})--(\ref{fitSG}),
evolved up to the $W$ mass.

We impose cuts on the lepton that are appropriate for
measurements \cite{CDF} at the CDF experiment.
A cut of 20 GeV was imposed on the $E_{T}$ of the emitted lepton so that the
integration region for $E_{T}$ is
$20\,{\rm GeV}\le E_{T} \le {1\over 2} M_{W}$.
The parameter $x_{{\mathbb P}}$ was integrated over the range
$M_{W}^{2}/s \le x_{{\mathbb P}} \le 0.01$.

\begin{figure}
\centerline{
\psfig{file=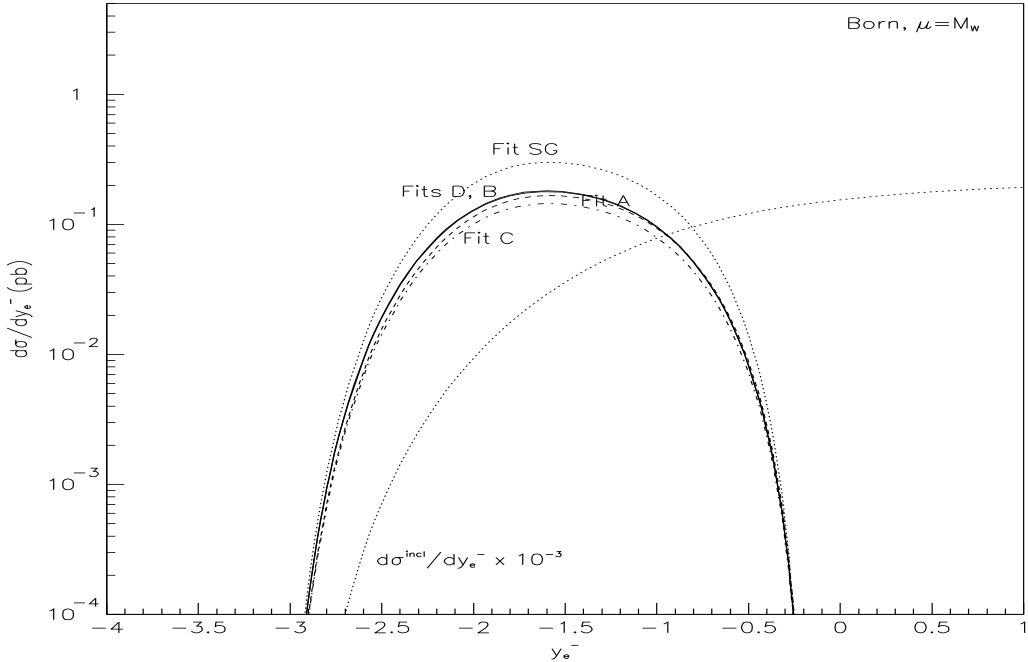,height=4in,width=6in,clip=}}
\caption{\sf Rapidity Distribution of $e^{-}$ in $W^{-}$ production.
        A cut $x_{\mathbb P} < 0.01$ was used.  }
\label{fig1Wm}
\end{figure}

Fig.~\ref{fig1Wm} shows our results for $W^{-}$ production.
The diffractive cross sections using fits B and D differ only by
a few percent and are represented
by solid curves which overlap, while those using fits A and C are denoted
by the dashed and dot-dashed curves, respectively.
These curves exhibit a strong fall-off
in the region $y_{e}>-0.25$ that is a consequence of the requirement
of a rapidity gap.
The overlap between the two solid curves suggests that, at least for this
particular process in this kinematic region, the extra soft term in fit
D for the quark in Eq.~(\ref{fitD}) does not contribute much.
The lower dotted curve is the inclusive cross section rescaled by $10^{-3}$.

The diffractive cross section is about 3\% to 4.4\% of the nondiffractive
one at the leftmost-edge of the plots (at $y=-3$) depending on the fit used.
The cross sections using fit B or fit D are about 45\% larger
than those using fits A and C at $y=-3$, with the ratio
decreasing as $y$ increases.
Since we only include the
$q{\bar q}\to W$ channel in the calculations, and the quark densities are
nearly the same for the five fits because of the small contribution of the
soft term, this difference is mostly due to the
effect of evolution.  The presence of a large amount of glue in
fits B and D increases the evolved quark densities relative to those in
fits A and C.
This tendency is even more pronounced for the SG fit (upper dotted
curve), where the
glue is concentrated at large $\beta $.

\begin{figure}
\centerline{
\psfig{file=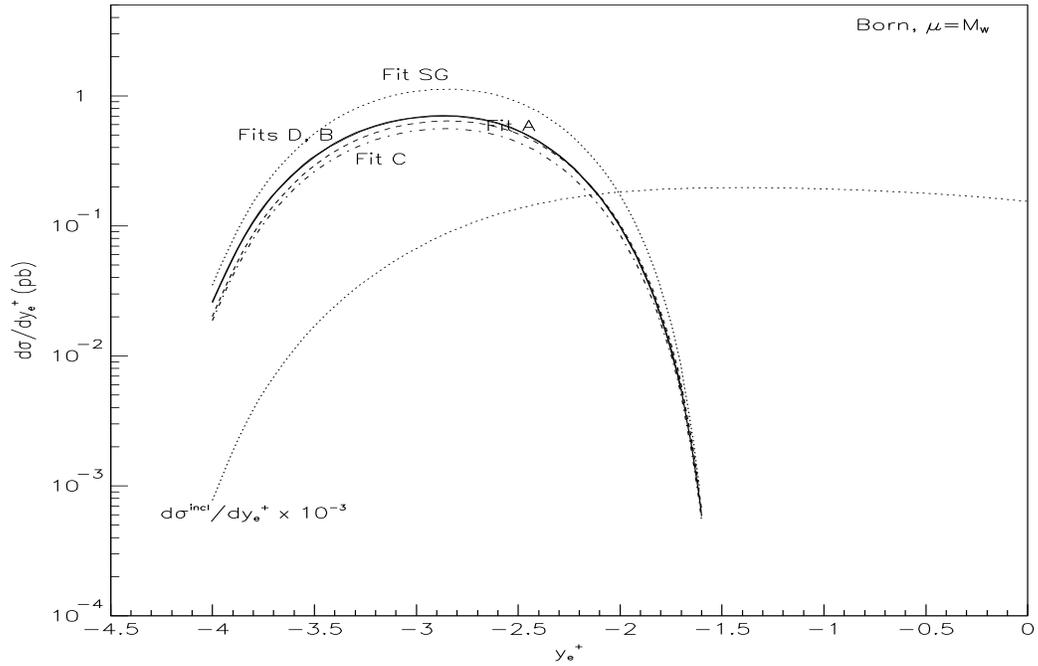,height=4in,width=6in,clip=}}
\caption{\sf Rapidity Distribution of $e^{+}$ in $W^{+}$ production.
        A cut $x_{\mathbb P} < 0.01$ was used.  }
\label{fig1Wp}
\end{figure}

The corresponding cross sections for $W^{+}$ are shown in
Fig.~\ref{fig1Wp}.
The cross sections are larger than for the $W^{-}$, because a valence
up quark from the proton can be used to make a $W^{+}$, especially at
large negative rapidities.
In the plot,
the rapidity gap exists for $y_{e^{+}}>-1.6$.  As in the case of $W^{-}$
production, the diffractive cross section using fit B or D
is larger than the one with fit A or C, by about 35\% at
$y_{e^{+}}=-4$, decreasing with $y_{e^{+}}$.

%-----------------------
\subsection{Comparison to CDF data for $W$ production}

The CDF collaboration has presented preliminary data on diffractive
$W$ production from $p{\bar p}$ collisions at ${\sqrt s}=1800 \, \rm{GeV}$
\cite{CDF,CDF.ANL}.  The $W's$ are produced with a rapidity gap in the region
$2.4 < |\eta | < 4.2$.
They find that
the fraction of diffractive to
non-diffractive
$W$ production
is \cite{CDF.ANL}
$R_{W}=[1.15 \pm 0.51({\it stat}) \pm 0.23 ({\it syst})]\%$.
% Comment (LA):   This is from the talk given by Bagdasarov
% (from Jim's notes of the
% ANL mtg).
% From ref. 8,  $R_W=[2.0 \pm 1.9(\it{stat} \oplus \it{syst})]\%$.

In Table \ref{table:cdfWrg} we present our diffractive fractions
using Eq.~(\ref{vbincldfcs})
with $x_{\mathbb P}^{max}=0.017$, which we
determined from the CDF plot \cite{CDF.Pot} relating $\eta _{\it max}$
and $x_{\mathbb P}$.
The fractions are about a factor of two smaller than the measured rates
but are all within the experimental uncertainties.  They are
computed with the diffracted hadron being allowed to be either
the proton or the antiproton.

\begin{table}
\begin{center}
    \begin{tabular}{||c|c|c|c|c||}
        \hline
                & Fit A & Fit B & Fit C & Fit D \\
        \hline
        $W^{+}+W^{-}$ & 0.56\% & 0.67\% & 0.50\% & 0.66\% \\
        \hline
    \end{tabular}
\end{center}
\caption{\sf Diffractive fractions using CDF cuts.}
\label{table:cdfWrg}
\end{table}

%======================================================
\section{Diffractive Jets}
\label{sec:jet.calcs}

In this section, we present our results for jet production.
We imposed the following cuts on the
jet cross sections.  These represent the effect of appropriate
experimental cuts \cite{CDF,D0.1}
and of cuts to improve the significance of the signal.
\begin{itemize}

\item
    We require that two jets are produced in the same half of the
    detector, i.e., $y_{1}y_{2}>0$, where $y_{i}$ is the rapidity of jet
    $i$.  This eliminates the region where the jets are in
    opposite hemispheres, since that region is well populated by
    non-diffractive events but is relatively unpopulated by
    diffractive events, because of the rapidity gap requirement.

\item
    Each jet is required to have a transverse energy $E_{T}$ greater
    than 20 GeV.  This ensures that we are definitely in the
    perturbative region for the jets, but the cut could be
    relaxed.

\item
    Each jet's rapidity satisfies $|y|>y_{cut}\equiv 1.8$.

\end{itemize}
Next, we integrated over the rapidity of
one of the jets
to obtain a single jet distribution, but still subject to the
above cuts on the other jet.  Eqs.~(\ref{diffcs1})
and (\ref{nondiffcs1}) were used for the diffractive and inclusive
cross sections, respectively, with the parton distributions
evolved to the scale $E_{T}$.
The integration limits used for the $E_{T}$ integral were
$E_{T}^{\rm min}=20\, {\rm GeV}$ and
$E_{T}^{\rm max}=\sqrt {s}/(e^{-y}+e^{y})$,
while the $x_{{\mathbb P}}$ integral was performed up to
$x_{{\mathbb P}}=0.01$.
In the following discussion, we will denote the rapidity of the final state
jet by $y_{\rm jet}$ instead of $y$.

\begin{figure}
\centerline{
   \psfig{file=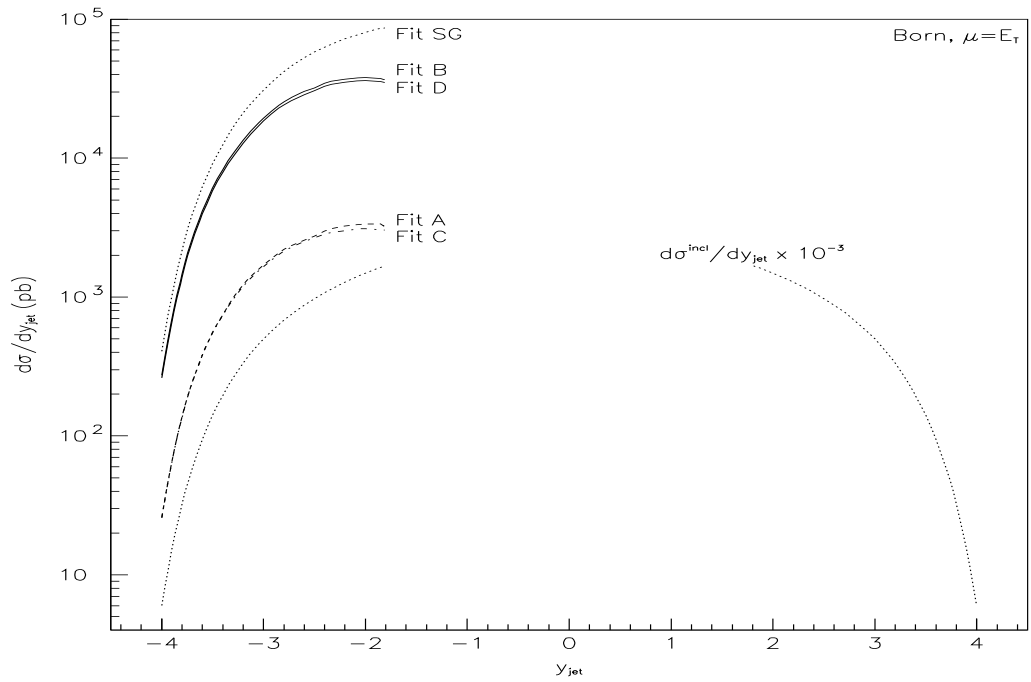,height=4in,width=6in,clip=}}
\caption{\sf Rapidity Distribution of Jet Cross Sections.}
\label{fig.jet}
\end{figure}

The resulting cross sections are shown in Fig.~\ref{fig.jet}.  There are no
points in the
middle part of the plot because of the rapidity cut used, as described above.
The lower (upper) solid curve, which results from using
fit D (fit B), is about 10 times larger than both the dashed and
dot-dashed curves, in which fits A and C, respectively, were used.
This reflects the sensitivity of this
particular type of cross section to the gluon content of the pomeron.
The lower dotted curve, which is symmetric about $y=0$,
represents the inclusive cross section
scaled down by a factor of $10^{-3}$.

We also show in Fig.~\ref{fig.jet} the cross section obtained
when the pomeron parton density with a singular gluon is used
(fit SG), the upper dotted curve. The resulting cross section
is about $1.5$ to $2.5$ times larger than that obtained using fit B or D.

\begin{figure}
\centerline{
   \psfig{file=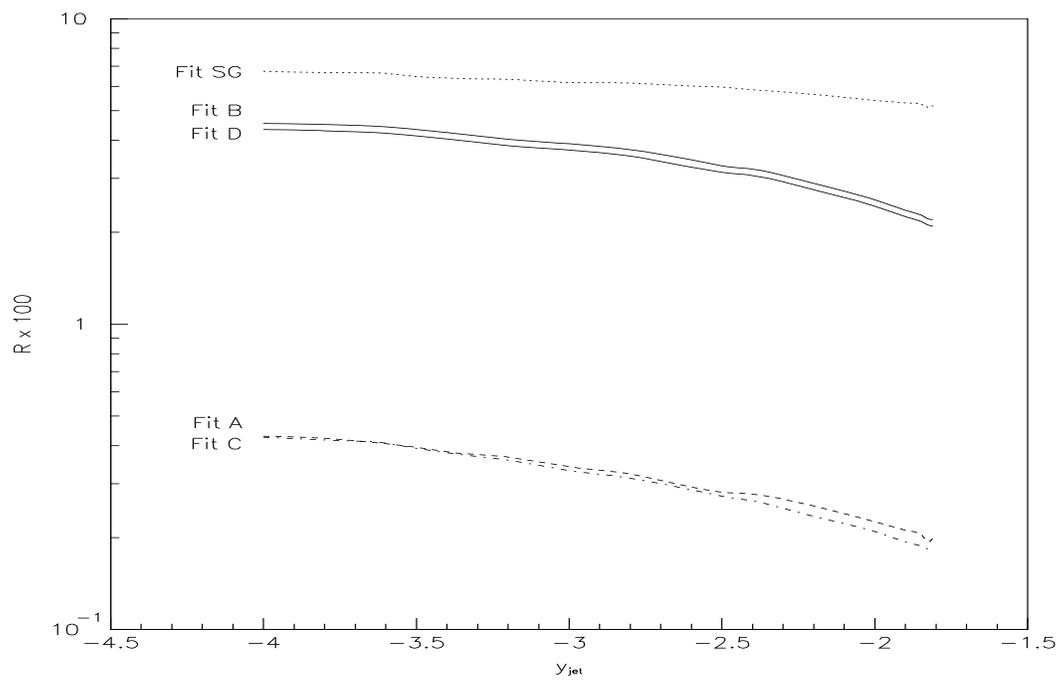,height=4in,width=6in,clip=}}
\caption{\sf Diffractive Jet Production Rates.}
\label{fig.jetrates}
\end{figure}

The diffractive jet fractions are shown in Fig.~\ref{fig.jetrates},
where $R \times 100$ is plotted as a function of $y_{\rm jet}$, with
$R={d\sigma ^{\rm jet, diff} /dy_{\rm jet}\over
   d\sigma ^{\rm jet, incl}/dy_{\rm jet}}$.
As in
Fig.~\ref{fig.jet}, the solid curves correspond to the rates when
fits B and D are used,
while the dashed and dot-dashed curves denote the rates for
fits A and C, respectively.
For the non-singular distributions,
one finds that the rates $R$ are largest when fit B is used, varying
from 4.5\% to 2.2\%.  With fit D, which differs from fit B by the absence
of a soft term in the quark distribution, the rates are
about tenths of a percent
lower. The rates obtained with fits A and C are almost identical and
range from 0.43\% to about 0.18\%.
The rates are largest at $y_{\rm jet}=-4$ then decrease as $y_{\rm jet}$
increases.
Of course, the large rates for the distributions B and D,
both with the large glue,
directly result from the fact that there is a gluon induced subprocess.
For comparison, we also show the even higher
rates obtained using the singular gluon fit SG (dotted curve),
which vary from 6.7\% to 5.2\%.

We end this section by making comparisons with preliminary data
on diffractive dijet
production from CDF and D0 at ${\sqrt s}=1800 \, {\rm GeV}$.
CDF has measured dijet data both with a rapidity gap
requirement \cite{CDF} and with Roman pots \cite{CDF.Pot}
along the antiproton beam direction.
In the first
case, the cross section for dijets produced opposite a
rapidity $(\eta )$ gap in the region $2.4<|\eta |<4.2$ is measured.
Each jet is required
to have a minimum $E_{T}$ of 20 GeV and rapidity $|\eta | > 1.8$.
They also measure the
dijet cross section without a rapidity gap, i.e., what we refer to
in this paper as the
{\it inclusive} cross section.  The diffractive fraction they
measure is \cite{CDF.ANL}
$R_{JJ}=[0.62 \pm 0.04 \pm 0.09]\%$.  The fractions that we obtain
using the above cuts
are shown in Table \ref{table:cdfJJrg}.  The rates obtained with
fit D or B are about nine
times larger while those obtained with fit C or A are about 25\%
smaller than the
measured value.
Our calculation assumes that either the antiproton or the proton
is diffracted.
% Comments (LA):  I used x_pom^max=0.017
% (corresponding to the gap 2.4< |\eta|<4.2.

\begin{table}
\begin{center}
    \begin{tabular}{||c|c|c|c|c||}
        \hline
                & Fit A & Fit B & Fit C & Fit D \\
        \hline
        ${\sigma ^{\rm jet,diff}\over \sigma ^{\rm jet,incl}}$
        & 0.47\% & 5.6\% & 0.47\% & 5.4\% \\
        \hline
    \end{tabular}
\end{center}
\caption{\sf Diffractive fractions for dijet production using
         CDF cuts (rapidity gap).}
\label{table:cdfJJrg}
\end{table}

With their Roman pot triggered diffractive sample, CDF
has measured a diffractive
fraction of $R_{JJ}=[0.109 \pm 0.003 \pm 0.016]\%$.
The data in this sample correspond
to $x_{\mathbb P}$ in the range $0.05<x_{\mathbb P}<0.1$, with the jets having
minimum $E_{T}$ of 10 GeV.  The fractions we obtain using the
same kinematic cuts
are presented in Table \ref{table:cdfJJpots}.  The ones obtained with
fits D, B are an order
of magnitude larger than the data,
while those obtained with fits C, A are about
3--4 times larger.
Our calculation assumes that only the antiproton is diffracted.
Since our diffractive parton densities are fitted to data in
which the proton may be excited, our predictions should be
reduced by about $(15\pm 10)\%$ \cite{ZEUSdata} before being
compared with data in which the isolated proton is detected.
However, even after this reduction, our predictions are still well above the
measured diffractive fraction.

\begin{table}
\begin{center}
    \begin{tabular}{||c|c|c|c|c||}
        \hline
                & Fit A & Fit B & Fit C & Fit D \\
        \hline
        ${\sigma ^{\rm jet,diff}\over \sigma ^{\rm jet,incl}}$
        & 0.33\% & 4.45\% & 0.37\% & 4.24\% \\
        \hline
    \end{tabular}
\end{center}
\caption{\sf Diffractive fractions for dijet production using
         CDF cuts (Roman pots).}
\label{table:cdfJJpots}
\end{table}

Finally, D0 also has some preliminary data \cite{D0.1} on diffractive
dijet production.  They require a
rapidity gap opposite the dijets which
have $E_{T}^{min}=12 \, {\rm GeV}$ and
$|\eta _{jet}| > 1.6$.  The diffractive fraction they measure
is $R_{JJ}=[0.67 \pm 0.05]\%$.  Our
calculated fractions are shown in Table \ref{table:d0rg}.
Again, the fractions obtained using
fits D, B are an order of magnitude larger than the data,
while those obtained
with fits C, A are
consistent with the data.
Our calculation assumes that either the antiproton or the proton
is diffracted.

\begin{table}
\begin{center}
    \begin{tabular}{||c|c|c|c|c||}
        \hline
                & Fit A & Fit B & Fit C & Fit D \\
        \hline
        ${\sigma ^{\rm jet,diff}\over \sigma ^{\rm jet,incl}}$
        & 0.8\% & 10.4\% & 0.9\% & 10.0\% \\
        \hline
    \end{tabular}
\end{center}
\caption{\sf Diffractive fractions for dijet production using
         D0 cuts (rapidity gap).}
\label{table:d0rg}
\end{table}

%========================================
\section{Conclusions}
\label{sec:concl}

We have presented
fits to the DIS diffractive structure functions measured by the ZEUS
collaboration, together with
a lowest order calculation of
the resulting
diffractive cross sections for
vector boson production and jet production from
$p{\bar p}$ interactions at the Tevatron.
Since we
used pomeron parton distributions
fitted to data on diffractive
DIS, the rates represent a realistic prediction of the cross
sections,
{\em given the assumption of factorization}.

The quark distributions are fairly well determined, at least in
overall size, and we see that diffractive $W$ and $Z$ production
is predicted to be a few percent of the inclusive cross section,
at least if suitable cuts are made to select kinematic
configurations that are preferentially populated in diffractive
events.

We have derived fractional rates $R = \sigma ^{\rm diff}/\sigma ^{\rm incl}$
which are much smaller than
those obtained in the benchmark studies \cite{BI}
of vector boson production.
The rates we have calculated are 0.08\% (0.03\%) for $W\ (Z)$
production using fit D, and a cut $x_{\mathbb P} < 0.01$.
For the case of lepton decays of vector bosons, we are able to
increase these fractions to several percent by
requiring the lepton to have large rapidity, on the side opposite
the rapidity gap.
Substantial deviations from these rates
would indicate a breakdown of hard scattering factorization for
diffractive processes.
In fact, the preliminary data from the Tevatron appear to be
consistent with our calculations, within large errors.

The wide range of gluon distributions that is permitted
by the DIS data yields
predictions for the diffractive jet cross section
at the Tevatron
(see
Fig.~\ref{fig.jet}) which differ by an order of
magnitude.  Given the assumption of factorization,
the smaller jet cross sections, from fits A and C,
represent the lowest reasonable cross sections since they
correspond to a pomeron without glue at $Q_{0}^{2} = 4 \, {\rm GeV}^{2}$.
However, the fits with the large gluon initial distributions
(fits B, D, and SG) are preferred for the ZEUS DIS data,
in agreement with the
results \cite{H1} from the H1 collaboration.

For jet production, rates of 4.5\% to 2.1\% in the rapidity range
$-4<y_{\rm jet}<-1.8$ were calculated using fit D or B
(with the large glue).
The corresponding rates for
the calculations using fit A or C
(with the small gluon distribution),
in the same rapidity range, vary from
0.43\% to 0.18\%.
We are working on using data \cite{ZEUSphoto} on diffractive
photoproduction of jets to reduce greatly the uncertainty in the
gluon density.  This will improve the accuracy of our
predictions, particularly for diffractive jet production.
We can already see from the results reported by the ZEUS
collaboration \cite{ZEUSphoto} that parton distributions with a
substantial amount of glue are preferred.

As we saw at the end of Sect.\ \ref{sec:jet.calcs}, only the
lowest of the jet cross sections we compute for the Tevatron is
consistent with the data.  These are the cross sections for which
there is no initial glue in the parton densities. In the case of
the Roman pot data, our calculated cross sections are all well
above the experimental value; these are the data for which the cuts
can be most directly
implemented
in our calculations. These results
combined with the ZEUS measurement of a substantial gluon
component in the pomeron already are suggestive of a breakdown of
factorization.

With a modest increase in accuracy,
comparison with actual data for diffractive $W$, $Z$ and jet
cross sections will permit a good test of the factorization
hypothesis.
As explained in Ref.\ \cite{nonfact}, the results of
such a test will be important as a probe of the space-time
structure of hadron-hadron scattering at high energy.

%\newpage
\section*{Acknowledgments}

This work was supported in part by the U.S.\ Department of Energy
under grant number DE-FG02-90ER-40577, and by the U.S. National
Science Foundation.
We are grateful for many discussions with our colleagues,
particularly those on the CTEQ and ZEUS collaborations and with
M. Albrow, A. Brandt, and J. Dainton.

\end{document}